# Influence of personal values and the adoption of analytical tools using laddering methodology

Igor Barahona Torres*

Department of Statistics,
Mathematics and Computer Science,
Cátedras CONACYT,
Autonomous University Chapingo,
56230, Km. 37.5 Highway
México-Texcoco, México
Email: jihbarahonato@conacyt.mx
*Corresponding author

Alex Riba

Statistical Services Centre,
University of Reading,
Harry Pitt Building,
Whiteknights Road,
Reading RG6 6FN, UK
Email: a.ribacivil@reading.ac.uk

James Freeman

Decision and Cognitive Sciences,
Manchester Business School,
University of Manchester,
Manchester M15 6PB, UK
Email: jim.freeman@mbs.ac.uk

**Abstract:** Analytical tools in business management are understood as a combination of information technologies and quantitative methods used to assist stakeholders to make better decisions. The contemporary business environment is dramatically changing by the massive accumulation of data. Now, as never before, the use of analytical tools must be expanded to take advantage of this growing digital universe. This article will apply the laddering technique to see how personal values (or managerial functions) influence a company's adoption of analytical tools. A set of ten in-depth interviews are conducted with CEOs, analytics consultants, academics and businessmen in order to establish quantitative relations among attributes, consequences and personal values. Two 'easy-to-read' outputs are provided to interpret our results. The most important links are quantitatively associated through an implication matrix, and then visually represented on a hierarchical value map. Guidelines for improving the use of analytical tools are provided in the last section.

## 1 Introduction

The emergence of more powerful computers, broadband Internet and smart-phones together with the proliferation of internet devices in such items as TVs, cars, refrigerators and more, have generated massive amounts of data, of a magnitude never before seen in the history of humankind. We live in data-driven societies. According to Turner et al. (2014) the digital data generated worldwide is now growing and will continue to grow by 40% each year and into the next decade. This will expand to include not only the increasing number of persons connected to the internet, but also the number of connected devices. For instance, from 2013 to 2020, this accumulation of data will grow by a factor of 10; that is, from 4.4 trillion gigabytes to 44 trillion. The authors also estimated that around 22% of all accumulated data in 2013 was suitable for analysis through the use of analytical tools. This is in contrast with the 35% which is expected to be reached by 2020, mostly due to the growth of data generated from embedded systems that are focused on managing real time applications. We find it interesting that the owners of nearly 80% of this ‘digital universe’ will be private companies, like Google, Twitter and Facebook. In this respect, the countries’ governments will be required to introduce regulations, in order to successfully cope with such issues as privacy, security and copyrights.

From this perspective, it is clear that traditional decision-making approaches, which are fundamentally based on subjective information, are becoming insufficient for dealing with this increasing complexity. Senior managers should dip into different analytical approaches to make the most of the data available. The greatest challenge during the next 10 years will be to find new approaches on how to extract relevant information from the data available in order to make better decisions. Competitive advantages will be given to organisations that are capable of extracting information from the data available and that will make better decisions based on this quantitative evidence. Several factors could influence managers on whether or not to adopt new analytical tools. The degree of access to financial, technical and human resources are critical variables when deciding whether to adopt new analytical tools. The extent of support received by senior managers is another critical variable. Personal values might also be considered as drivers for expanding the use of analytical tools.

In terms of literature about the adoption of analytical tools, it is a fact that most research is focused on investigating its quantitative and structured aspects; that is, the quantitative dimensions of the analytical tools as the amount of access to financial, technological and financial resources are widely discussed by Hoerl and Snee (2010), Anderson-Cook et al. (2012), Deming (2000), Steinberg et al. (2008), Davenport and Harris (2007), Gardner (2004), CMMI (2013), Davenport, Harris and Morrison (2010) and others. The impact of such qualitative variables as the type of leadership, managerial functions or personal values has received less attention from researchers. Consequently, the investigation of how personal values impact the adoption of new analytical tools represents a novel contribution to the field of business analytics. Given this, the main objective of this paper is to quantify how personal values (or managerial functions) influence the adoption of new analytical tools. To answer this, our article is structured in six sections. First, we present a review of the literature regarding values and analytical tools, followed by the methodology and our research objectives. The results, which were obtained through our laddering analysis are explained in section four; and then a set of practical guidelines is presented for stakeholders who are interested in expanding the use of analytical tools. And finally, the last section presents a discussion.

## 2 Personal values and the adoption of analytical tools

Personal values have been investigated since ancient times. Raz et al. (2003) documents the teachings of Socrates and Aristotle which included some basic definitions of personal values and their relation to human behaviour. Rokeach (1973) widely discusses the role of personal values as a critical factor of the human being, influencing ones decisions, choices and evaluations. Rokeach defines personal values in the following way:

> "*To say that a person has a value is to say that he has an enduring prescriptive or proscriptive belief that a specific mode of behaviour or end-state of existence is preferred to an opposite mode of behaviour or end-state. This belief transcends attitudes toward objects and toward situations; it is a standard that guides and determines action, attitudes toward objects and situations, ideology, presentations of self to others, evaluations, judgments, justifications, comparisons of self with others, and attempts to influence others. Values serve as adjustive, ego-defensive, knowledge, and self-actualising functions*" (Rokeach, 1973, p. 25).

England (1967), on the other hand, investigates connections amongst personal values, social norms and emotions for the purpose of revealing the impact of values on decision making. Personal values are a priority and the first option when a decision is made by the individual. Consider the following example: if a decision has to be made regarding the use of analytical tools, a particular senior manager will review his or her personal values as the first option. Later, she or he will gather all the information required for reaching a '*well-matched*' decision with his or her personal values. Hemingway and Maclagan (2004) states that personal values are a sort of heuristic device for making decisions. Personal values may function as a determinant factor when deciding about corporate strategy, the vision statement or even the adoption of new analytical tools. In this case, if the values of a particular senior manager include honesty, integrity and commitment, then the essence of those personal values will influence the corporate strategy and the mission statement at the organisational level.

As for the empirical research which investigates the role of personal values on business management, Johnson, Melin and Whittington (2003) conducted a study to find out how daily operations and personal values impact a company's strategic planning process. von Krogh, Ichijo and Nonaka (2000) described the relationship between the knowledge-creation process and senior manager values. Hoerl and Snee (2010) affirm that there is a direct tie between the level of adoption of an analytical tool and the perceived degree of its usefulness. Senior managers first perceive that analytical tools add value to the organisation, and then they decide to adopt them. The higher the perception of an analytical tool as useful, the greater the probability that it will be adopted. From this perspective, the question then is: What type of managerial functions have a positive influence on a manager's perception of the usefulness of analytical tools? To answer this we must use quantitative research methods that can uncover hidden and unstructured aspects of personal values. In addition to this, research on how personal values may influence the level of adoption of analytical tools for better decision making is also required. Our research methods are explained in detail in the following sections.

## 3 The methodology

### *3.1 The laddering technique*

Application of laddering methodology to the field of business management was first proposed by Reynolds and Gutman (1988). From the very beginning, it proved to be a powerful tool for eliciting personal values in different areas such as marketing, management and research surveys, among others. This methodology offers several benefits such as, for example, it allows analysing relations between the values of two or more individuals and the variable of interest, which in our case is the level of adoption of analytical tools. Three specific objectives were formulated for this research, as follows:

- conduct 10 in-depth interviews with senior managers, consultants and academics, in order to obtain information about their personal values and the level of adoption of analytical tools
- apply the laddering methodology to identify those personal values (or managerial functions) that influence managers in their adoption of new analytical tools
- provide practical guidelines to stakeholders who are interested in expanding the use of analytical tools.

Herrmann, Huber and Braunstein (2000) reported that the laddering technique was first developed by psychologists during the 1960s, as a tool to investigate patients' values. The technique offered several advantages in comparison with other interviewing techniques; such as, it is relatively simple to implement and understand, and its capacity to generate '*easy to read*' results. Consequently, the laddering technique was rapidly adapted to other areas, such as marketing, human resources and business management. The laddering technique was incorporated in the discipline of marketing by Reynolds and Gutman (1984). During this adaptation, the means-end theory, which describes the linkages between personal values and behaviour, was also incorporated. According to Reynolds and Gutman (1988), the term '*laddering*' within the context of market research refers to an in-depth, one-on-one interview technique, which is applied to understand how customers transform attributes of any given product or service into meaningful associations with respect to self by following the means-end theory.

That is to say, information about products or services is given within a certain context, then the individual forms a conception on the degree of suitability (the means) which is able to complete that product or service, in order to satisfy a specific personal need (the end). Basically, it is possible to apply the means-end theory to describe the way the customers give consequences and assign importance to any given product or service. As stated by Reynolds and Gutman (1988), the context generally affects the importance assigned by the customer. For example, in times of economic crisis, customers who see their incomes reduced may consider avoiding going to clubs and casinos or buying luxury goods. Figure 1 offers a visual representation of the means-end model.

Gutman (1982) states that values related to enjoyment, social recognition and good health play a decisive role in attaching importance to their respective consequences. For instance, the value '*social recognition*' is related to '*financial independence*'; thus, the latter will have attached a consequence rated as important. But how can personal values and their consequences be uncovered? A novel contribution of this research is use of the laddering technique to find values that are relevant to the adoption of new analytical tools by an organisation. This research is also pertinent given the dearth of documented cases in literature regarding use of the laddering technique in the field of business analytics.

The operationalisation of laddering proposed by Reynolds and Gutman (1988) includes a format that uses a series of directed questions such as '*Why is that important to you?*', where the ultimate objective is to uncover the connections between the key perceptual elements of *attributes* (A), *consequences* (C) and *values* (V). In this way, different levels of abstraction, represented by the assembled ladders, provide a deeper understanding of how the manager's perception of the adoption of new analytical tools is processed from what could be called an emotional perspective. Figure 2 illustrates three levels of abstraction with the form *attributes* (A), *consequences* (C) and *values* (V).

**Figure 1** The means-end model (see online version for colours)

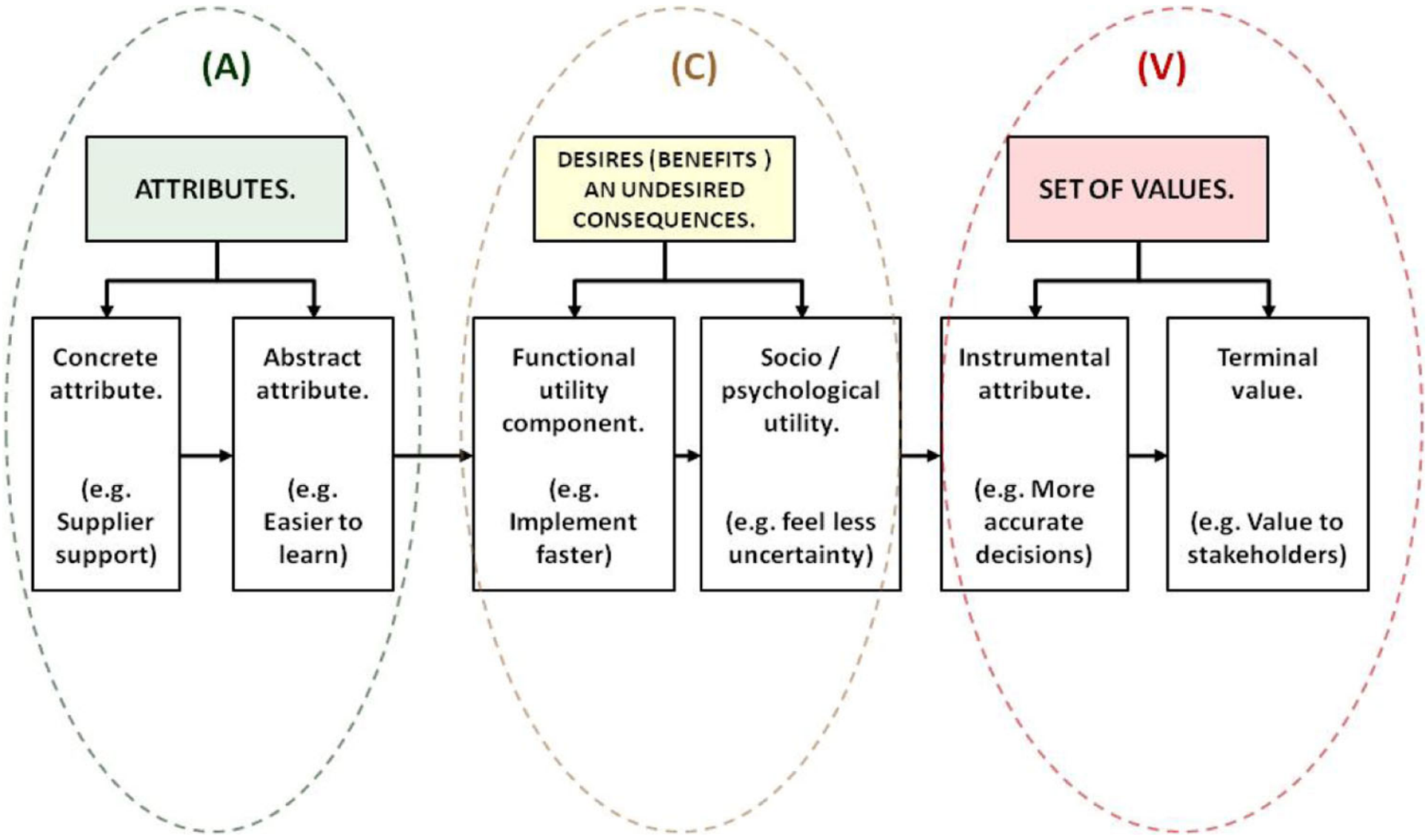

*Source:* Adapted from Herrmann and Braunstein (2000)

**Figure 2** Example of ladder constructed with data collected during our in-depth interviews (see online version for colours)

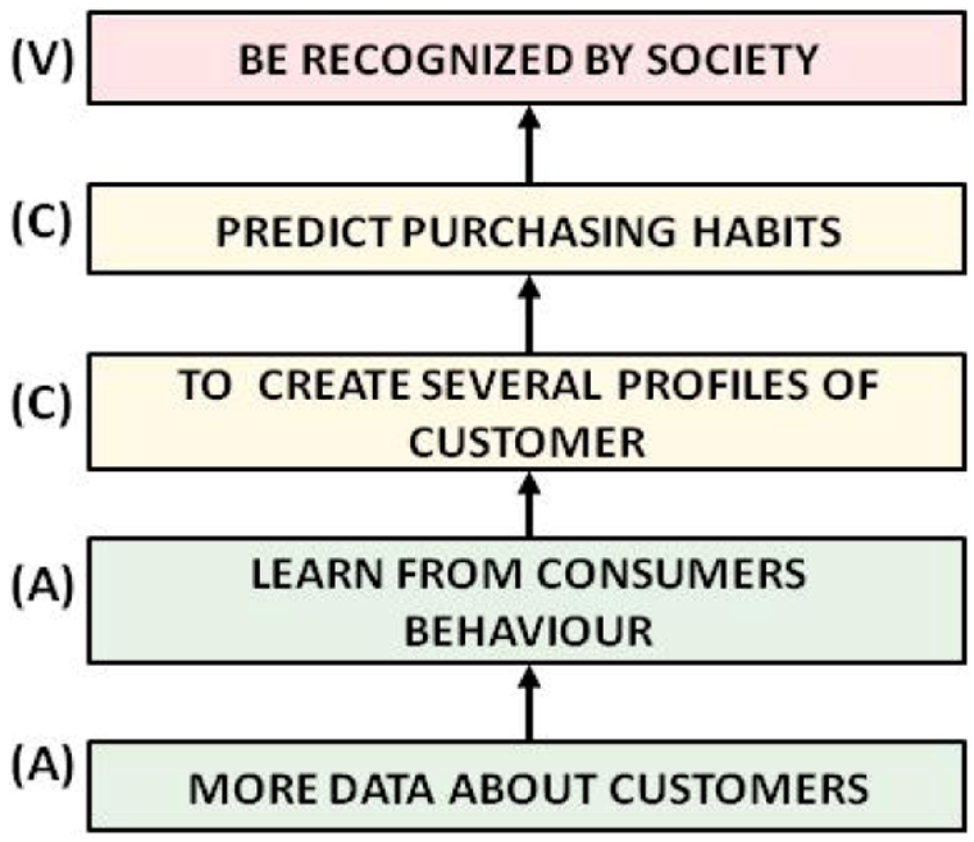

An important feature of the laddering technique is the capacity to cause the responder to think critically about his or her personal motivations, by gradually moving him or her forward to higher levels of abstractions; beginning with the attributes and finishing with the personal values. Once the eliciting process is finished, data collected from multiple responders can be summarised in what is called the implication matrix (IM). This matrix is relevant because it allows us to identify the most dominant connections with the form (A)→(C)→(V). It is also an input for other analysis instruments. The hierarchical value map (HVM), which is assembled based on the implication matrix, is a type of cognitive map with all the connections graphically represented. In this research, the HVM interpretation allows us to understand which personal values are relevant when deciding to adopt new analytical tools. Another HVM contribution to this research is its ability to differentiate the most relevant managerial functions for the adoption of new analytical tools, not only by attributes but also by disclosing how the adoption of these new analytical tools in itself delivers higher consequences, and finally, its relevance in terms of personal values. This understanding of personal values might be used as the foundation for developing strategies to adapt data-driven strategies.

### *3.2 In-depth interviews through the laddering technique*

According to Reynolds and Gutman (1988), some conditions are indispensable in the interview environment in order to obtain valuable data. First, a friendly atmosphere must be created to make the responder feel confident and willing to be introspective, to look inside and search for feelings and motivations. It is advisable that the interviewer provide some introductory comments, stating that there are no right or wrong answers. This will make the responder feel relaxed. The interviewer should insist that the main purpose of the in-depth interview is to talk about perceptions, feelings and notions, completely free of judgments. The responder should be viewed as an expert on the topic in discussion. The interviewer should always keep in mind that the ultimate objective of the interview is to understand the way in which the responder, based on his feelings and motivations, perceives the use of analytical tools. It is also important that the interviewer act merely as facilitator of this self-discovery process and all personal opinions and judgments must be avoided. The strategy suggested by Reynolds and Gutman (1988) is to begin with the questions that may seem obvious, very simple or even stupid so that the respondent hopefully feels more confident and more willing to talk. It is also advisable that the respondent speak most of the time and the interviewer remain silent.

Table 1 presents six methods for eliciting the responder's emotions and feelings. The reader may find it difficult to distinguish one from another, since they are pretty similar. According to Herrmann, Huber and Braunstein (2000), an effective and in-depth laddering interview should include the combination of all the methods presented. Plenty of experience and knowledge from the interviewer is required in order to intelligently apply any specific method to any specific individual. For Reynolds and Gutman (1988), the key is: the more familiar the interviewer is with the methods presented in Table 1, the better he or she will be able to manage, combine and integrate them. The main topic throughout the interview must be the person (not the service, the idea, the concept or the product).

**Table 1** Common interview techniques used in laddering

| *Laddering method* | *Description* | *Example questions* |
|---|---|---|
| 1. Evoking any given situational context | It is feasible to reach a ladder when the respondent thinks about one past moment when he/she interacted with the service, product or concept. | *When was the last time you applied an analytical tool? Why it was important to you?* |
| 2. Supposing the absence | Another method to reach a ladder is by asking for feelings and sensations, but given the hypothetical situation when there is a lack of the service, product or concept. | *How do you make a decision if you could not access to computers and analytical tools? Why this (absence or presence) is important to you?* |
| 3. Negative – inverse laddering | Sometimes the responder is unable to articulate his/her feelings. If this is the case, a negative question may help to clarify responder's mind | *Given that situation. What would happen if you don't use an analytical tool?* |
| 4. Back in time | Invite responder to go backward in time is another method to elicit feelings and motivations. | *Do you know how your former boss used to apply analytical tools to reach a decision? Can briefly explain?* |
| 5. Third person experience | Sometimes the responder will find it difficult to talk about her/his experience. In this case, evoking a third person will stimulate the responder to speak about his/her own experience. | *What problems are your colleagues struggling due to the lack of use in analytical tools? Why do you think that is important for your colleague?* |
| 6. Redirecting methods: silence, rapport and check | Silence on one part of the interview will be helpful to maintain the responder thinking about feelings and motivations. Likewise all the types of interviews the checking and rapport are important. | *Rapport in the interviews occurs when both (respondent and interviewer) feel they are in sync and relate each other. Interview rapport should include mutual attention, positivity and coordination.* |

*Source:* Adapted from Reynolds and Gutman (1988)

According to Reynolds and Gutman (1988), the typical in-depth interview should last between 60 to 75 minutes. Approximately five to ten ladders can be assembled from each one. In this research, a total of ten in-depth interviews were carried out and 84 different ladders assembled.

## 4 Results

Three types of professionals were interviewed: consultants, academics and businessmen. All were asked if the interview could be recorded and no one refused. Consequently, the same number of digital records is available for anyone who requests them. After reviewing the records carefully, the first analysis consisted of classifying all the ideas presented in groups according to its closest meaning. For example, the concept "*data should support more accurate decision making*" is close in meaning to "*data accessibility supports better decision making*" and they were classified in the same group. Similarly

"*online data facilitates communication*" was found to be related to "*sharing information is easier when data is online*"; consequently they were classified in the same group entitled '*online data*'. By listening to our digital records six times each, a total of 34 groups were created. The next step was to classify them into the three basic groups: (A)→(C)→(V). Figure 3 shows the 34 groups of attributes, consequences and values. In addition, the frequency for each group is shown in the right column.

**Figure 3** Summary of contents of the in-depth interviews

| | Attributes | Count |
|---|---|---|
| 1 | Data is accessible and supports decisions | 17 |
| 2 | Data online | 12 |
| 3 | Goal setting | 11 |
| 4 | Standardized procedures | 10 |
| 5 | High skilled staff | 7 |
| 6 | Enough support | 6 |
| 7 | High tech | 6 |
| 8 | Communication with customers and suppliers | 5 |
| 9 | Creativity to propose new ideas | 5 |
| 10 | Information outside the organization | 5 |
| 11 | Market research | 5 |
| 12 | The most efficient structure | 5 |
| 13 | Flexibility on management | 4 |

| | Consequences | Count |
|---|---|---|
| 16 | Analyse data from market | 5 |
| 17 | Continuous learning | 3 |
| 18 | Distinctive competence | 6 |
| 19 | Exceeding the customer expectations | 7 |
| 20 | Good image of the organization | 6 |
| 21 | Improve data analysis | 18 |
| 22 | Improving process | 7 |
| 23 | Improving results | 14 |
| 24 | Knowledge of data | 7 |
| 25 | Long term relationships with actors | 7 |
| 26 | Lower cost | 5 |
| 27 | More money | 13 |
| 28 | Staff efficiency and motivation | 11 |

| | Values | Count |
|---|---|---|
| 29 | Add value to stake holders | 13 |
| 30 | Being a leader | 9 |
| 31 | Communication and trust | 9 |
| 32 | Honesty and credibility | 5 |
| 33 | Passion, Quality and Excellence | 11 |
| 34 | Serving the society | 12 |

Data from the in-depth interviews was processed and analysed, but always from the perspective of how it relates to the main objective of the study: to understand the relationships between the elements. Grunert and Grunert (1995) state that a distinctive feature that separates the laddering methodology from other in-depth interview techniques is its capacity for '*crossing over*' the qualitative nature of the in-depth interview by generating quantitative and structured data as the final output. By creating a scoring matrix from the ladders it is possible to quantitatively identify dominant pathways and connections among the key elements. As we will show later, the implication matrix displays the number of times that one concept leads to another. This is a square matrix which shows size by the number of concepts that we are representing. If the relationship among 34 concepts is analysed, then the matrix will be on the order of 34 × 34. Based on Reynolds and Gutman (1988), the implication matrix may include two types of relationships: direct and indirect. Consider the following case as a better explanation of direct relationships.

In Figure 4 you will note that A→B (autonomy for using data **leads to**→ stimulate creativity and innovation) is a direct relationship. In the same way, B→C (stimulate creativity and innovation **leads to** → use more data for decision making), C→D (use more data for decision making **leads to** → achieve goals and objectives), and D→E (achieve goals and objectives **leads to** → being recognized for society), are all direct relationships. Consequently, direct relationships from the attributes to values are elicited, and finally a typical ladder is completed. On the other hand, continuing with the example of Figure 4, you will notice that A→C, A→E and C→E are indirect relationships.

**Figure 4** Direct relationships in a typical ladder (see online version for colours)

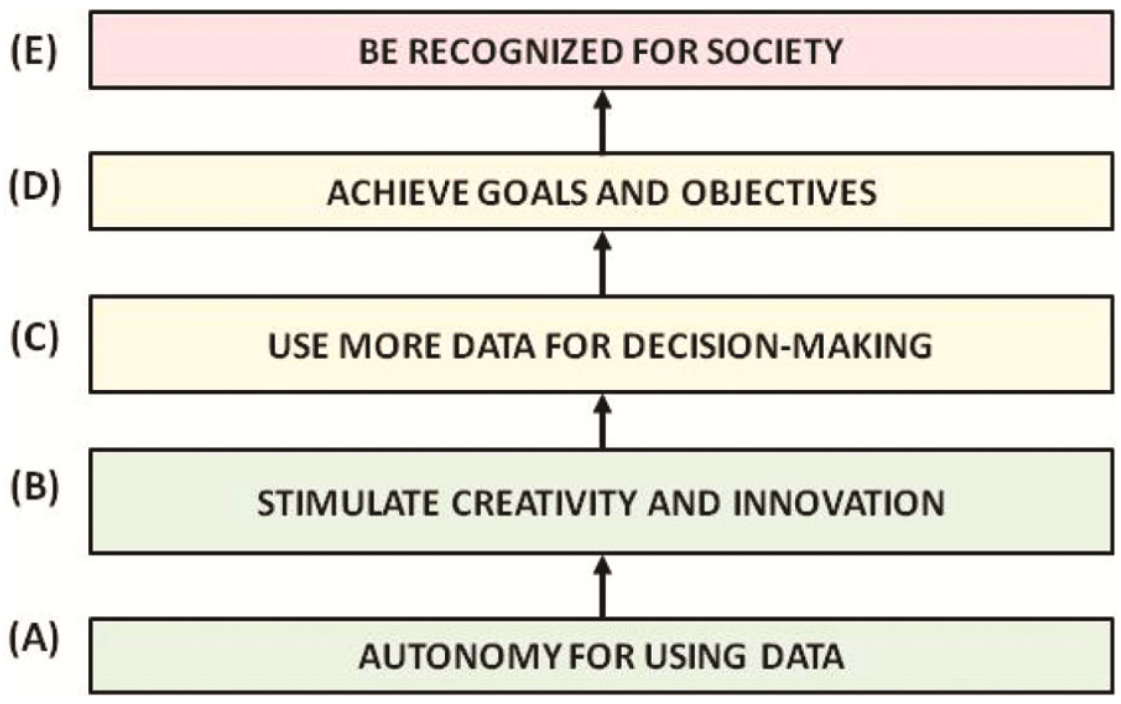

Table 2 presents the implication matrix. For each *row-column* presented, the frequency indicates the number of times that a *row-element* directly and indirectly leads to a *column-element*. Numbers are separated by the symbol '–', which is used to differentiate direct from indirect relationships. While the numbers on the left represent direct relationships, those on the right are indirect relationships. For instance, the first attribute "*data is accessible and support decisions*" (row 1) directly leads to "*improve data analysis*" (row 21) 17 times, and only once indirectly (17:01). Similarly, "*staff efficiency and motivation*" (row 28) leads directly to "*communication and trust*" (row 31) eleven times and zero indirectly (11:00).

After computing the elements in the implication matrix, now we have to calculate the totals for both direct and indirect relationships in order to identify the most dominant relationships. The consequence '*improve data analysis*' (row 21) has the biggest number of direct connections with the row '*add value to stakeholders*', (29) equal to 18 relationships. The next step is to represent all the identified connections in one '*easy-to-read*' visual diagram. Reynolds and Gutman (1988) define the HVM as "*A way to graphically represent the most dominant connections. It is a visualisation of the linkages across levels of abstraction, starting with attributes and finishing with values*". Based on Henneberg et al. (2009), a common approach would be to include all the connections which are comprised of at least four or more direct relations. Thus, 84 ladders that meet this requirement are represented on the HVM.

**Table 2** Implication matrix for the adoption of new analytical tools (see online version for colours)

| *Element* | *16* | *17* | *18* | *19* | *20* | *21* | *22* | *23* | *24* | *25* | *26* | *27* | *28* | *29* | *30* | *31* | *32* | *33* | *34* |
|---|---|---|---|---|---|---|---|---|---|---|---|---|---|---|---|---|---|---|---|
| A 1 Data is accessible and supports decisions | | | | | 8-00 | 17-01 | 4-00 | 4-00 | 4-00 | 0-03 | 0-05 | 0-09 | 0-08 | 0-13 | 0-09 | 0-09 | 0-08 | 0-11 | 0-04 |
| A 2 Data online | 3-00 | | 3-03 | | | 7-03 | 4-00 | 4-00 | 5-00 | | 0-06 | | | 0-06 | | | | | 0-04 |
| A 3 Goal setting | 3-00 | 3-00 | 2-00 | 4-00 | 2-00 | 5-00 | 2-00 | 2-04 | 13-00 | | 5-05 | | 4-00 | 0-06 | 0-06 | 0-05 | 0-10 | 0-11 | 0-12 |
| A 4 Standardised procedures | 2-00 | 1-00 | 1-00 | 3-00 | | 6-00 | 3-00 | 0-03 | | | 0-06 | 0-03 | 11-06 | 0-05 | 0-10 | 0-10 | 0-05 | | |
| A 5 High skilled staff | 3-00 | | 3-00 | 1-00 | | 5-00 | 2-00 | | 0-03 | | | 0-04 | 4-00 | | | 0-10 | 0-08 | 0-10 | 0-06 |
| A 6 Enough support | | | 0-03 | | | 6-00 | 3-00 | | | 0-06 | | | 4-00 | 0-03 | 0-08 | 0-09 | 0-05 | 0-09 | 0-06 |
| A 7 Hightech | 0-02 | 0-03 | | 0-04 | | 3-00 | | | 0-03 | | 0-07 | 0-03 | | | 0-10 | | | | |
| A 8 Communication with customers and suppliers | 5-03 | 4-00 | 3-00 | 2-00 | | | 4-00 | 0-08 | 0-02 | | | | 0-05 | 0-05 | | 0-08 | 0-10 | 0-02 | 0-05 |
| A 9 Creativity to propose new ideas | | 4-00 | 5-00 | 5-08 | | | 4-00 | 0-08 | 8-04 | | 0-06 | 1-08 | 5-00 | 0-06 | | 0-08 | 0-10 | 0-07 | 0-08 |
| A 10 Information outside the organisation | 7-04 | | 2-00 | 1-00 | 3-00 | | | | | 0-05 | | | | | 0-06 | | | | 0-08 |
| A 11 Market research | 4-00 | 24K) | 0-04 | 0-04 | | | | 0-02 | 0-06 | 0-04 | | | | | 0-04 | | | | |
| A 12 The most efficient structure | | | | | | | | | | | 5430 | 3-00 | 6-05 | 0-06 | 0-06 | 0-06 | 0-04 | 0-02 | |
| A 13 Flexibility on management | | 2-00 | 4-00 | 5-00 | 0-05 | 2-02 | | | | | | 0-04 | | 0-01 | | 0-06 | | | 0-06 |
| A 14 Respond more quickly | | 2-00 | 3-00 | 3-07 | | | 3-00 | 0-03 | 6-00 | 0-02 | 0-05 | 0-04 | 3-00 | | 0-05 | 0-08 | | 0-04 | 0-05 |
| A 15 To innovate products and services | | | 4-00 | 3-00 | | 0-10 | | | 3-00 | 0-02 | 0-02 | 0-05 | 4-00 | 2-00 | | 0-02 | | 0-04 | 0-05 |
| C 16 Analyse data from market | | 10-00 | 0-05 | 3-00 | | | 0-03 | 0-05 | 6-00 | | | 0-06 | | 0-05 | | | | | |
| C 17 Continuous learning | | | 4-00 | | | 5-00 | 0-03 | | | 0-02 | | | 0-06 | 0-04 | | 0-04 | 0-03 | 0-06 | 0-06 |
| C 18 Distinctive competence | | | | 3-00 | | | | | | | 6-00 | 0-08 | | 0-06 | | | | 0-06 | 0-06 |
| C 19 Exceeding the customer expectations | | 4-00 | | | 0-04 | | | 3-00 | | | | | 2-00 | | | | | | |
| C 20 Good image of the organization | | | | | | 0-02 | | | | 8-00 | | | 0-02 | | | | 0-06 | 0-04 | 0-05 |
| C 21 Improve data analysis | | | | | | | 6-00 | 5-00 | | | 04)6 | 0-05 | | 18-00 | 10-00 | | | 14-00 | |
| C 22 Improving process | | 11-00 | | | | | | | | 0-06 | 4-00 | 0-04 | 0-05 | 7-00 | 2-10 | 0-05 | 0-05 | 0-10 | 0-07 |
| C 23 Improving results | | 14-00 | | | | | | | | | | | | | | | | 0-10 | 0.07 |
| C 24 Knowledge of data | | | | | | | 4-00 | 7-00 | | | | | | | | | | | |
| C 25 Long term relationships with actors | | | | | | | | | | | | | | | 7-00 | | | | 0-08 |
| C 26 Lower cost | | | | | | | | | | | | 5-00 | | | | | | | |
| C 27 More money | | | | | | | | | | | | | | 13-00 | 12-00 | | | | |
| C 28 Staff efficiency and motivation | | | | | | | | | | | | | | 12-00 | | 11-00 | | 8-00 | |
| V 29 Add valueto stake holders | | | | | | | | | | | | | | | | | 6-00 | | |
| V 30 Being a leader | | | | | | | | | | | | | | | | | 6-00 | | |
| V 31 Communication and trust | | | | | | | | | | | | | | | | | | | 7-00 |
| V 32 Honesty and credibility | | | | | | | | | | | | | | | | | | | 14-00 |
| V 33 Passion, Quality and Excellence | | | | | | | | | | | | | | | | | 9-00 | | |
| V 34 Serving the society | | | | | | | | | | | | | | | | | | | |

Figure 5 presents the HVM. Element attributes, consequences and values (A→C→C) are assigned as follows: the lowest levels of abstraction or attributes are presented at the bottom of the HVM. In contrast, the highest levels of abstraction or values are found at the top of the map. Note that black arrows connect elements and indicate the direction of each relationship. The numbers in red represent how many times a given element leads to another.

**Figure 5** Hierarchical value map for the adoption of new analytical tools (see online version for colours)

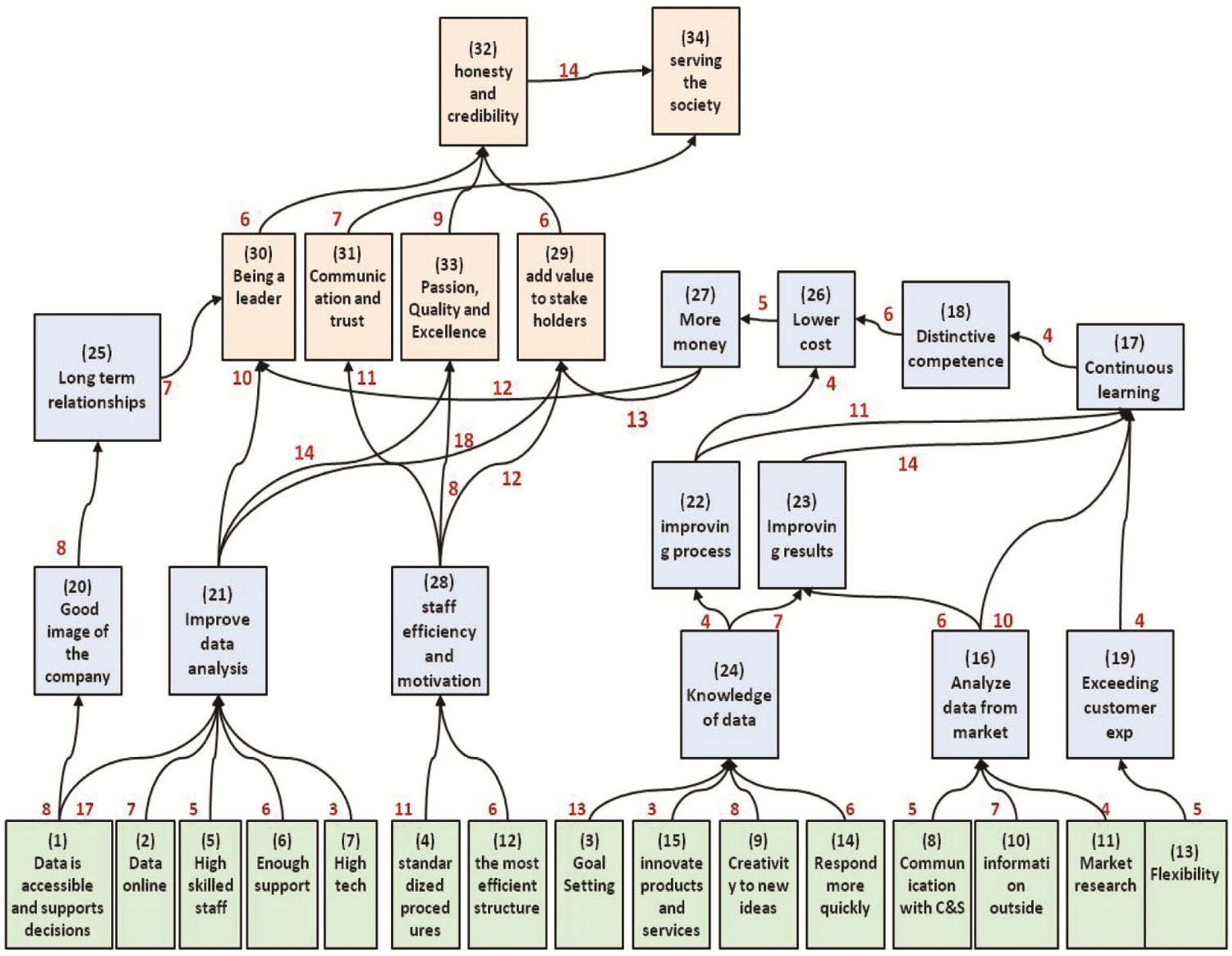

As shown in Figure 5, the attribute "*data is accessible and supports decisions*" **(1)** leads 17 times to the consequence '*improve data analysis*' **(21)**; likewise it leads 18 times to the value '*add value to stakeholders*' **(29)**. It also leads 6 times to the value '*honesty and credibility*' **(32),** and finally, this leads 14 times to '*service to the society*' **(34)**. Other elements are noteworthy for their high frequency in laddering to other elements. Namely, the attribute '*standardised procedures*' **(4)** leads 11 times to '*staff efficiency and motivation*' **(28)**, and 18 times to the element '*passion, quality and excellence*' **(33).** It also leads 9 times to the value '*honesty and credibility*' **(32)** which finally leads 14 times to the element '*serving the society*' **(34)**.

It is clear that the value which receives the highest number of relationships is also the most relevant. The attribute '*goal setting*' **(3)** leads 13 times to the consequence '*knowledge of data*' **(24)**, as well as 7 times to '*improving results*' **(23),** which leads 14 times to '*continuous learning*' **(17)**, and this 4 times to '*distinctive competence*' **(18),** which also leads six times to '*lower cost*' **(26)**. This leads five times to '*more money*'

**(27)**, which also leads 13 times to the value '*add value to stakeholders*' **(29)**, as well as six times to '*honesty and credibility*' **(32)**, and finally, 14 times to '*service to the society*' **(34)**. The cumulative frequency for this ladder is equal to 82. Therefore, there is quantitative evidence to show that there is a relationship between the attribute '*goal setting*' and the value '*serving the society*' which is equal to 82 direct connections (see Figure 6).

As mentioned before, our main goal is to disclose quantitative relationships between attributes and values, which at first glance may be hidden. Our starting point is the 10 attributes which have the largest number of relationships. These ten attributes represent more than 90% of the total relationships. A cumulative frequency following the order of (A)→(C)→(V) was calculated for each of them. Table 3 shows the results obtained.

**Figure 6** Frequency summary for the ladder starting with the attribute 'goal setting' (see online version for colours)

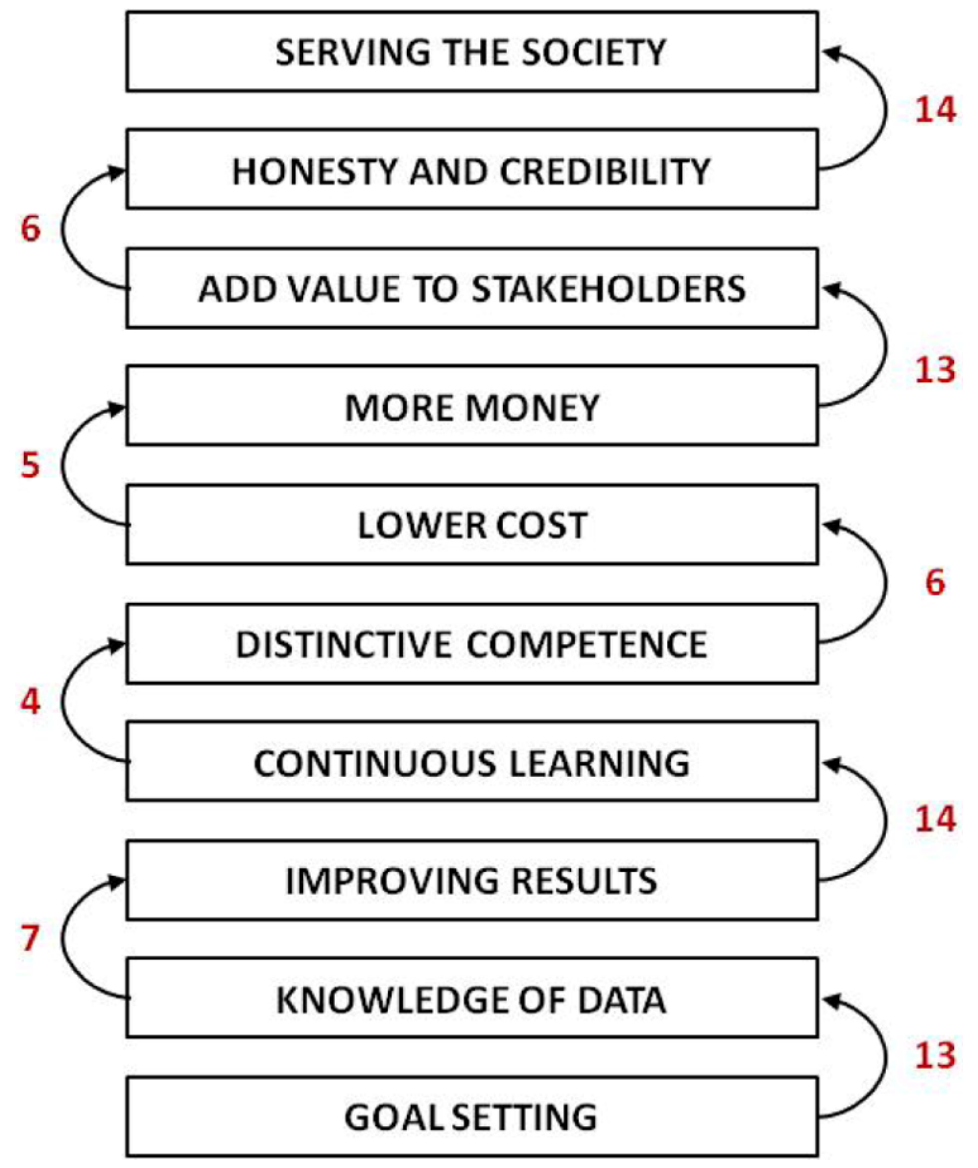

The attributes presented in Table 3 are divided into two groups. The first group gathers values related to personal achievement and individual effort; the second group shows values related to social skills and social interaction. The attribute '*goal setting*' had the biggest cumulative frequency, equal to 212, followed by '*creativity to propose new ideas*'. In the third position was '*information outside the organisation*'.

**Table 3** Table of frequencies for relations among attributes and values (see online version for colours)

| *Attributes* | *Personal achievement* | | | *Social values* | | | | | |
|---|---|---|---|---|---|---|---|---|---|
| | *Being a leader* | *Passion, quality and excellence* | *Total* | *Communication and trust* | *add value to stake holders and credibility* | *honesty and credibility* | *serving the Society* | *Total* | |
| **Goal setting** | 61 | | 61 | | 62 | 68 | 82 | 212 | **12%** |
| **Creativity to propose new ideas** | 56 | | 56 | | 57 | 63 | 77 | 197 | **11%** |
| **Information outside the organisation** | 54 | | 54 | | 55 | 61 | 75 | 191 | **11%** |
| Respond more quickly | 52 | | 52 | | 55 | 61 | 75 | 191 | **11%** |
| Communication with customers and suppliers | 42 | | 42 | | 43 | 49 | 63 | 155 | **9%** |
| Data is accessible and supports decisions | 27 | | 27 | | 45 | 51 | 47 | 143 | **8%** |
| Enough support | | 20 | 20 | | 24 | 30 | 44 | 98 | **6%** |
| Standardised procedures | | | 0 | | 23 | 29 | 43 | 95 | **5%** |
| Data online | | 21 | 21 | | | 30 | 44 | 74 | **4%** |
| The most efficient structure | | 14 | 14 | | | 23 | 37 | 60 | **3%** |
| **Total** | 292 | 55 | 347 | | 364 | 465 | 587 | 1416 | |

Given these results, social values are more influential when expanding the use of analytical tools than personal achievement values. While around 80% of the relationships ended with social values, 20% finished with personal achievement values. The social value '*service to the society*' had the highest cumulative frequency, equal to 587 or 33% of all relationships ending with this value. '*Honesty and credibility*' was in second place with a cumulative frequency equal to 465, representing 26% of the total relationships. In third place was '*added value to the stakeholders*', with a cumulative frequency of 364, representing 21% of total relationships. Note that these three social values concentrate 80% of total relationships. In fourth place we found the value '*being a leader*', with 292 direct relationships, representing 17% of all relationships. These four values concentrate 97% of the total connections.

With regard to attributes, '*goal setting*' got the highest number of connections, equal to 212 or 12% of the total. The '*creativity to propose new ideas*' was in second place with 197 connections, representing 11% of the total. In third position is the attribute '*information outside the organisation*' with 191 connections, representing 11% of the total. The attribute '*responding more quickly*' was ranked in fourth position with 191 connections. These four attributes concentrate 45% of the total outgoing relationships.

## 5 Practical guidelines to stakeholders

Three attributes draw our attention: '*goal setting*', '*creativity to propose new ideas*' and '*getting information from outside*'. On the other hand, we find that another three managerial functions (or personal values) represent 97% of the total relationships; these are: '*service to the society*', '*honesty*' and '*adding value to the stakeholders*'. At this point, we find that the initial question of how values influence the adoption of new analytical tools is partially answered. In order to fully answer it, the discussion now centres on providing guidelines to stakeholders who are interested in expanding the use of analytical tools.

With respect to the attribute '*goal setting*', which is the most influential attribute, Thompson and McEwen (1958) stated that it is almost impossible for an organisation to last indefinitely if goals are not formulated and deployed. The first suggestion provided here is to formulate and deploy pertinent goals by following well-proven processes. Shalley (1995) documents best practices to formulate '*high impact*' goals. Some of the important features of '*high-impact*' goals are also discussed by this author; among them, the capacity to guarantee the long term survival of the company and respond properly to changes in the environment. Locke et al. (1990) suggest that there is a direct relationship between goal setting and productivity. Goal setting increases productivity when individuals accept and commit to specific challenging goals and receive appropriate feedback. The famous article written by Locke et al. (1981) offers plenty of help to strengthen the goal-setting process. This resource is widely suggested for all our readers.

Amabile et al. (1996) defines creativity, the second most influential attribute, as the production of novel and useful ideas on any domain. Creativity is the starting point for innovation and thus for experimentation, which is strategic to expand the use of analytical tools. Creative and critical thinking are indispensable in order to generate helpful new ideas. However, critical thinking might also act as a killer of creativity by discouraging those who propose new ways to solve problems or to create things. For instance, phrases such as '*Yes, we tried this*', "*No, it won't work*' and '*This doesn't make any sense*' discourage the generation of new ideas. That is: “*Before criticising the idea of another, you must first find two valid reasons to support it.*” This ratio of two supportive versus one critical view will reinforce the growth of creative thinking.

The last discussed attribute refers to the capacity of ‘*getting information from outside*’. According to Roome (1992), the external environment includes all actors outside the organisation, such as suppliers, customers, society, governments and economic conditions, etc. It is extremely important to constantly monitor the external environment for signals of change in the markets which may require the modification of assumptions used in the models, metrics or methodologies. In this case, we suggest the monitoring process proposed by Ruff (2006) as this classifies efforts to monitor products and services first, and later, the macro-environment formed by markets, industries, and economic and political factors. Management tools such as FODA, Hoshin Kanri, the external perspectives of the balanced scorecard, and the five competitive forces of Porter, among others, are strongly advised to successfully monitor the external environment.

Based on the frequencies found in Table 3, the most influential value (or managerial function) is ‘*serving the society*’. Managers should, in this respect, make sure that the four dimensions proposed by Carroll and Buchholtz (2011) are present in their organisations. The first dimension embraces the financial returns that an organisation is supposed to achieve. In the second place, the organisation should be able to follow all laws and regulations. Ethical aspects refer to compliance with moral codes and proper conduct, which are expected to be fully respected. Finally, philanthropic responsibility is understood as the expected behaviour which society finds positive as it promotes and reinforces healthy habits. The more an organisation grows in these four dimensions, the more it serves society.

The second value discussed here is ‘*honesty*’. Becker (1998) defines this as “*the refusal to pretend that facts of reality are other than what they are*”. In other words, honesty is the appreciation of reality as it is. It is clear that honesty has a strategic importance for satisfying customer expectations and thereby achieves a competitive advantage. Levels of honesty are in direct proportion to the morals of the workforce and their commitment to the goal-setting process. A set of practical guidelines is provided to assist the manager with improving honesty. The first suggestion is to lead by example. By setting a moral and ethical standard, senior managers provide a solid reference of what is expected from everyone. Leaders should stress the importance of honesty to suppliers and customers by stating it clearly in the mission statement. Open and honest communication must be fomented at all levels of the organisation. This is an effective way to clear up rumours, negativity, or even dishonesty. All stakeholders should be encouraged to voice their concerns, ideas and even complaints without fear of defensive reactions. As for ‘*adding value to stakeholders*’, Porter (2008) widely discusses the role of the customers, employees, shareholders and society in order to create competitive advantages. In this respect, you must explicitly specify what benefits each stakeholder is receiving.

Figure 7 shows a double effect on the adoption of analytical tools. At the bottom of the organisational structure, the attributes ‘*goal setting*’, ‘*creativity to propose new ideas*’ and ‘*getting information from outside*’ influence daily operations, methods and operating processes of the organisation. At the middle level we find statistical engineering which establishes a link between *attributes-&-methods* and *values-&-strategy*. At the top of the structure the values ‘*serving the society*’, ‘*honesty*’ and ‘*adding value to the stakeholders*’ influence the mission and vision statements (*analytical thinking*).

**Figure 7** The influence of values and attributes on the adoption of new analytical tools (see online version for colours)

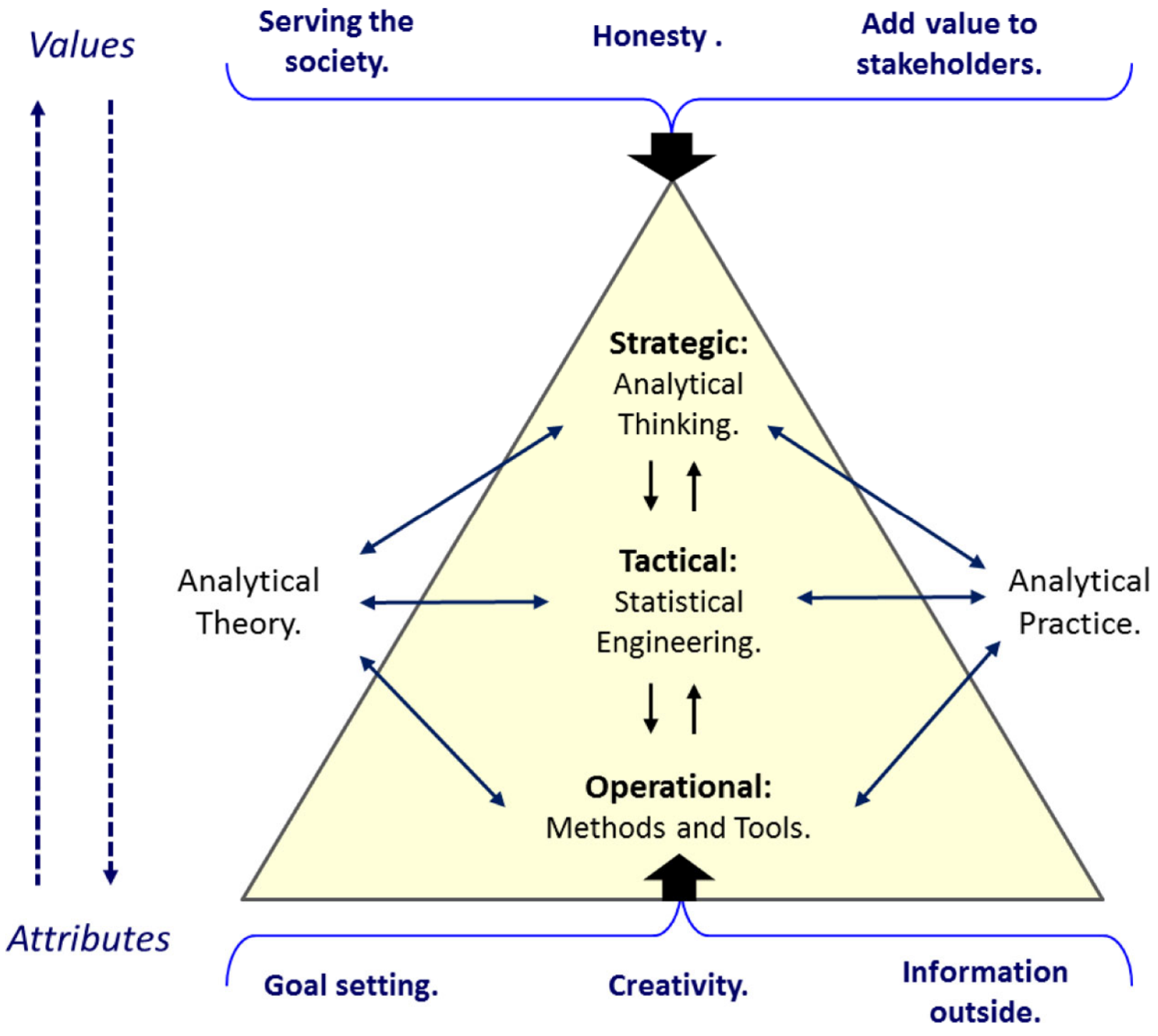

*Source*: Adapted from Hoerl and Snee (2010)

In short, initiatives to expand the adoption of analytical tools can be divided in two groups. The first focuses on influencing daily operations through an expanded use of analytical tools, namely:

- improve the goal-setting process
- stimulate creativity
- improve the quality and relevance of information from outside.

On the other hand, mission and vision statements should be based on solid values. Senior managers must be the main promoters of managerial functions such as:

- demonstrate by facts that the organization is ***adding value to the society***
- ensure that ***honesty*** is a big concern in the organisation
- ensure that ***expectations of the stakeholders*** are satisfied.

## 6 Discussion

In this research, the laddering technique is applied to investigate analytical tools and managerial functions. First, a general perspective of the contemporary business analytics environment is given. Then, formal definitions for personal values are provided from literature, within the context of management. A broad explanation of the laddering methodology is provided in section three, followed by the results obtained. Considering that we conducted only ten in-depth interviews, our results are intended as more illustrative than convincing. In other words, our contribution is more to demonstrate the versatility of the laddering technique and its capacity to cross over different disciplines rather than to throw compelling evidence on the adoption of analytical tools. Although results are limited in scope and relevance, a set of practical guidelines is provided to stakeholders interested in expanding the use of analytical tools in their organisations. Our guidelines are complemented by insights from other authors in order to make them more helpful and robust.

To this point, it is clear that additional research is needed to complement our results. A follow-up paper will be prepared, in which a second (and larger) set of managers will be asked for their '*managerial functions*' and the adoption of new analytical tools. We will propose that a quantitative model be prepared to test whether we can predict if analytical tools will be adopted or not. In this way, the complete framework will consider contributions in two ways: one methodological, implying a novel use of laddering to examine groups rather than consumers in order to solicit higher order beliefs of other kinds; and the second to reveal managerial perceptions of their own multidimensional roles and functions. While the first is provided here, the second will be the subject of a follow-up paper.

Here, we adopted the definition of standard operating procedure proposed by Walsh and Ungson (1991) in order to describe our attributes. It is evident that complementary research is pertinent to investigate what type of managerial functions (or personal values) would derive from different definitions of attributes; such as communication devices, assembly lines or office assistants. The point is to identify to what extent managerial functions are altered when a change is made to the concept of attributes. Similarly, more investigation is required on changes in the response variable as a consequence of interviewing different groups. For instance, how do the perceptions of academics, managers and consultants differ? This may produce other interesting results.

Since this longstanding technique has been widely applied to derive higher-order attributes for brands, its replication in the field of business analytics represents a novel contribution. At the time that this research was conducted, no documented cases of the use of this technique in the field of business analytics were found.